\documentclass[conference]{IEEEtran}
\IEEEoverridecommandlockouts

\usepackage[utf8]{inputenc}

\usepackage{amssymb,amsmath}
\usepackage{float}
\usepackage{tikz}

\def\BibTeX{{\rm B\kern-.05em{\sc i\kern-.025em b}\kern-.08em
    T\kern-.1667em\lower.7ex\hbox{E}\kern-.125emX}}

\title{Crowded MTC Random Access in NOMA XL-MIMO}

\author{
\IEEEauthorblockN{{Thiago Augusto Bruza Alves}}
\IEEEauthorblockA{\textit{Electrical Eng. Dept.}  - 
\textit{UEL}\\
Londrina, Brazil.\\
thiagobruza@outlook.com \vspace{-6mm}}

\and

\IEEEauthorblockN{{Taufik Abrão}}
\IEEEauthorblockA{\textit{Electrical Eng. Dept.} - 
\textit{UEL}\\
Londrina, Brazil.\\
taufik@uel.br \vspace{-6mm}} 

\thanks{This work was supported in part by the CAPES (Financial Code 001) and  National Council for Scientific and Technological Development (CNPq) of Brazil under Grant  310681/2019-7.}
}

\begin{document}

\maketitle

\begin{abstract}
Massive MIMO is one of the key technologies to support the growth of massive access attempts by devices, such as in {massive} machine type communication (mMTC). The evolution of antenna array technology brought the recent extra-large scale massive multiple input multiple output (XL-MIMO) systems, seen as a promising technology for providing very high-data rates in high-user density scenarios. Spatial non stationarities and visibility regions (VRs) occur across all huge XL array extension, since its large dimension is of the same order of the distances to the user equipment (UE). We investigate the random access (RA) problem in crowded XL-MIMO scenarios. The proposed {non-orthogonal multiple access} (NOMA) visible region extra large array (NVR-XL) protocol takes advantage of the power domain NOMA to allow access of two or more users colliding in the same XL sub-array (SA) selecting the same pilot sequence. The NVR-XL provides a reduction in the number of attempts to access the network, while improving the average sum-rate, as the number of SA increases. 
\end{abstract}	
\smallskip
\vspace{-4mm}
\section{Introduction}

The number of active devices keeps growing {jointly} with the number of access requests for different types of services (e.g. vehicles, sensors, mobiles, etc.), and in applications for 5G and beyond, as massive machine type communication (mMTC) and crowded mobile broadband (cMBB) use cases, the base station (BS) will need to attend simultaneously devices with a range of capabilities and deployments.
%

%

Power-domain NOMA {was} proposed initially to improve the spectral efficiency of wireless networks, sharing the same orthogonal resource (time and frequency), by superposition coding in transmitter side and successive interference cancellation (SIC) in receiver side \cite{Maraqa2020}.

{A vast number of devices trying to access the network at the same time is a medium access control challenge. Traditional random access (RA) schemes are not able to handle such a large number of requests \cite{Clazzer2019}. Pure RA schemes, like ALOHA, have severe performance limitation. The incorporation of NOMA significantly improves {the system} performance, {by} admitting two or more users per time slot {at same frequency} \cite{Silva22}.}

In a massive {RA} scenario, the number of user connection attempts far outnumbers the number of available pilot sequences. As a consequence, {the implementation of} a collision resolution protocols became essential to allow coherent communication. One well-known decentralized grant-based random access protocol for crowded massive multiple-input multiple-output (MIMO) systems is the strongest user collision resolution (SUCRe), which takes advantage of MIMO properties \cite{Emil2017}, giving preference to users with good channel conditions.

%
%
Combining NOMA and SUCRe was an important evolution. In \cite{Pereira2021} compare to the classic SUCRe scheme, the proposed NOMA-RA protocol allows users that try to access the medium with the same pilot signal to resolve collisions by distinguishing them in the power domain and attaining a superior sum-rate performance with reduced average latency.




{In \cite{Nishimura2020}, an adaptation of SUCRe in extra large scenarios is proposed.} 
%
The concept of VR is explored in XL-MIMO context, since there is a relationship between the geographic area and the portion of the array that is visible from that area.
When a certain UE is present in that region, the corresponding visibility region is accessible; on the other hand, when UE exits the VR, it encounters a new set of clusters. The probability that SA $b$ is visible by UE $k$ can be explored to improve the RA.


The contribution of this paper is twofold: \textbf{a}) {explore the} overlapping VR in the { proposed 
NVR-XL RA} protocol{, by combining} the advantages of XL-MIMO {scheme} and NOMA, allowing two users to select the same pilot {sequence at the same} SA. {The superimposed signal is processed on a SA basis, which is performed SIC to decode each user's information.} {\textbf{b}) carry out a comprehensive numerical simulation analysis by comparing the proposed NVR-XL scheme with the literature SUCRe-XL protocol by deploying three different metrics.}

\vspace{-2mm}

\section{System Model} \label{System_model}

\begin{figure}[htbp!]
\vspace{-3mm}
    \centering
    \includegraphics[width=.99\linewidth]{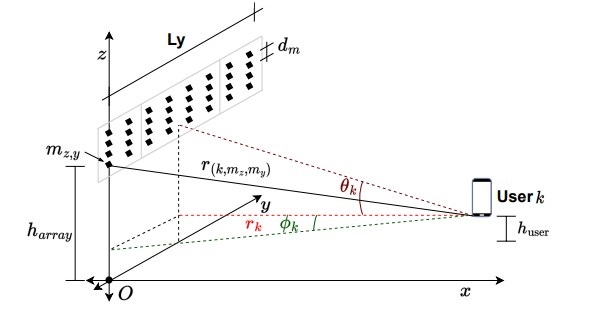}
    \vspace{-7mm}
    \caption{Diagram of the $k$th user at the URA XL-MIMO with an impinging sphere waveform from elevation angle $\theta$ and azimuth angle $\phi$. The $M = M_y \times M_z$ are horizontally and vertically equally spaced with spacing $d_m$. }
    \label{fig:system}
\end{figure}

We consider an XL-MIMO setup where the BS deploys an extra-large antenna array, operating in a time-division-duplexing (TDD). Without loss of generality, we assume a uniform rectangular array (URA) is placed on the $y$-$z$ plane with a total of $M$ antenna elements, i.e., $M_y \times M_z$, where $M_y$ and $M_z$ denote the number of antenna elements along the $y$- and $z$-axis, respectively, with the firth line of an array starting at the origin $O$ and deployed along the ordinate axis, as in Fig. \ref{fig:system}.
Similarly to \cite{Nishimura2020,Marinello2022}, we consider a simplified bipartite graph model in XL-MIMO, the array is divided into $B$ subarrays (SAs), {each composed of} a fixed number of $M_b = M/B$ antennas, ensuring the minimum antennas elements $\left (M_b \geq 50 \right)$, to achieve the inherent properties of massive MIMO.

All single-antenna user devices are randomly distributed and grouped in set $\mathcal{U}$, $\mathcal{A} \subset \mathcal{U}$ be the subset of active UEs, {with temporarily} dedicated data pilots, and $\mathcal{K} = \mathcal{U} \setminus \mathcal{A}$ be the set of inactives UEs (iUEs). 
The devices in set $\mathcal{K}$ do not have dedicated {pilots; if they} want to be active, they must be {assigned one}. Hence, the number of iUEs {in the cell} is $|\mathcal{K}| = K$. The BS only allocates pilots to active devices and reclaims the pilots when needed. The iUEs make a RA attempt with probability $P_a$.   

Let $\mathcal{M}$ denote the set of all BS SAs and $\mathcal{V}_k \subset \mathcal{M}$ the subset of SAs visible to the $k$-th UE. To compose the mathematical modeling, we assume that the subset $\mathcal{V}_k$ is generated at random, where each SA is visible with probability $P_b$ by UE $k$, like in \cite{Nishimura2020,Marinello2022}. This probability simulates the influence of random obstacles and scatterers in the environment {interacting} with the signals transmitted by/to the UEs, resulting in VRs. The existence of VRs as evidenced by measurements in \cite{Carvalho2020}.

We consider a multi-user communication system based on time-frequency coherence blocks and inactive UEs have RA blocks to realize RA attempts. We define $\tau_{\text{RA}}$ as the number of RA mutually orthogonal pilot sequences that are possible $\textbf{s}_1,\dots,\textbf{s}_{\tau_{\text{RA}}} \in \mathbb{C}^{\tau_{\text{RA}} \times} 1$, each with length $\tau_{\text{RA}}$, and $\left \| \textbf{s}_{\tau_{\text{RA}}} \right \| = \sqrt{\tau_{\text{RA}}}$.

\begin{figure}[htbp!]
\vspace{-5mm}
\centering
\includegraphics[width=.95\linewidth]{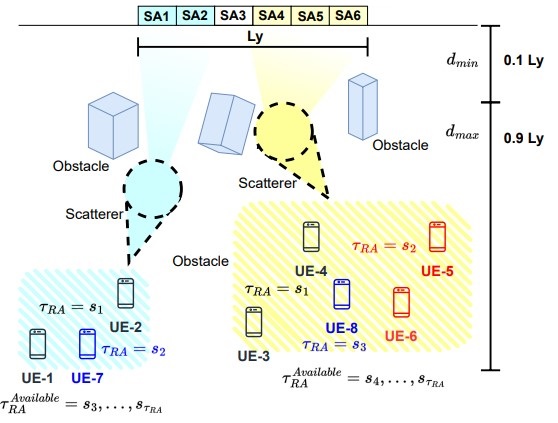}
 \vspace{-4mm}
\caption{Crowded MTC system with URA divided into a 6-SA; the colored regions in each SA define visible UEs, $P_b$ is visible success probability. }
\label{fig:distance}
\end{figure}

Since each SA visibility follows a Bernoulli distribution with success probability $P_b$, 
it is crucial to note that when an antenna is visible to a user, it indicates that the user may transmit/receive signals to/from this antenna with a non-zero channel gain \cite{Marinello2022}.
%
We assume {an overcrowded} scenario where $K$ iUEs $\left (K \gg M \right)$ are randomly distributed within a square cell with maximum distance $d_{max}$ and minimum distance $d_{min}$, respectively, as in Fig. \ref{fig:distance}.

\vspace{-5mm}
\subsection{Channel Model}

{In next, we model the} NLOS channel based on the position of each URA antenna element $m_{z,y}$, as in Fig \ref{fig:system}. 
The specific localization of $m_{z,y}$-th array element is represented by 3D vector $\textbf{w}_{m_{z,y}}$.
The physical dimension of the URA along the $y$- and $z$-axis are Ly $= (M_y)d_m$ and Lz $= (M_z)d_m$, where $d_m$ is the distance between elements.
The user-$k$ position is defined by 3D vector $\textbf{q}_k$, the distance of the $k$-th UE to the specific antenna element $m_{z,y}$ is determined by $r_{k,m_y,m_z} = \left \| \textbf{w}_{m_{z,y}} - \textbf{q}_k \right \|$ \cite{Lu2022}.
%
%
The large-scale fading includes path loss and shadowing, on the basis of an urban micro scenario model provided by: 
\begin{equation}
    \beta_{k,m_z,m_y} = 10^{-\kappa \log(r_{k,m_z,m_y})+\frac{g+\varphi}{10}},
\end{equation}

\noindent where $g$ = $-$34.53 dB is the path loss at the reference distance, the path loss exponent $\kappa$ = 3.8, and $\varphi \sim \mathcal{N}(0,\sigma^2_{\text{sf}})$ is the shadow fading, a log-normal random variable with standard deviation $\sigma_{\text{sf}}$ = 10 dB.

Similarly as \cite{Nishimura2020,Marinello2022}, we adopt the average large-scale fading for the $k$-th user at the $b$-th SA, $\beta_k^{(b)} =$ $\frac{1}{M_b} \sum_{m=1}^{M_b} \beta_{k,m}$. Let $\textbf{h}^{(b)}_k \in \mathbb{C}^{M_b \times 1}$ be the Rayleigh fading channel vector between UE $k \in \mathcal{K}$ and SA $b$, following $\textbf{h}_k^{(b)} \sim \mathcal{CN}$ $\left ( 0,\beta_k^{(b)} \textbf{I}_{M_b} \right )$.

\vspace{-3mm}
\section{Random Access for NOMA XL-MIMO}

\subsection{SUCRe protocol}

The SUCRe protocol \cite{Emil2017} is a distributed method {for resolving pilot collisions on} the user side. The protocol {consists of the following} steps: $\textbf{a)}$ $K P_a$ users randomly choose an uplink (UL) RA pilot and transmit it; $\textbf{b)}$ the BS receives such UE signals and responds with precoded downlink (DL) pilots; $\textbf{c)}$ Using the received signal, UEs can estimate the total of the signal strengths of UEs competing for the same RA pilot chosen by them, {compare such estimate to its signal gain}, and retransmit the chosen RA pilot if it judges it is the strongest UE among the contenders; $\textbf{d)}$ in conclusion, the BS allocates dedicated data pilots to the UEs not colliding in Step before. In mMTC scenarios $K$ is large when compared to the number of antennas $M$ $\left ( K \gg M \right )$, in contrast $P_a \ll 1$.


\subsection{Exploring VRs and NOMA in SUCRe Protocol}
The proposed NVR-XL protocol combines the {improvement} of non-overlapping VRs singular UEs along the SAs and overlapping VRs when two {users} can select the same pilot along the SAs. The Steps 1 and 2 are similar to the proposed in the SUCRe-XL protocol \cite{Nishimura2020}.

\noindent \textbf{Step I}: $KP_a$ UEs make an RA attempt. 

\noindent \textbf{Step II}: SAs send a precoded DL pilots to UEs from Step I. 

\noindent\textbf{Step III}: By exploring the power domain NOMA, users colliding pilots can be served at the same resource (spatial (SA), frequency, time, pilot). Due to imperfections in the SIC process, in this work, we accept collisions of up to two users, {\it i.e.}, performing up to one SIC step. However, in collisions between three or more users in at least one SA, the protocol fails to resolve the collision, and unsuccessful UEs are instructed to try again on the next RA with probability $P_{na}$, limited to a maximum of 10 attempts.

The collision resolution follows the distributed decision rule; if $\rho_k \sum_{m \in \mathcal{V}_k} \beta_k^{(m)} \tau_{\text{RA}} > \hat{\alpha}_{t,k} / 2 + \epsilon_k$ is true, the user decides to repeat its pilot transmission, otherwise, the user decides to remain inactive by pulling out the RA attempt and trying to communicate in next RA stage with probability $P_{na}$.

In this decision rule, the {\it bias} term $\epsilon_k$ is given by :
\vspace{-2mm}
\begin{equation}
\epsilon_k = \frac{\delta}{\sqrt{M_b} \cdot \sum_{b \in \mathcal{V}_k} \beta_k^{(b)}}, \vspace{-2mm}
\end{equation} where $\delta$ is a scale factor that may be adjusted in order to improve some system metrics. We adjust the scale factor $\delta$ numerically as in \cite{Pereira2021}, by using an exhaustive search approach so that it maximizes the sum-rate; hence, each $K$-iUE and $B$-SA configuration results in an optimal $\delta^{K,B^*}_{\textsc{nvr-xl}}$.

The main feature proposed in our NVR-XL protocol is the modification in the contention process of \cite{Nishimura2020}, wherein a pilot collision in the set $\mathcal{S}_t$ and occurring overlaps in VRs, the signal is processed in each SA to decode superimposed signals. Hence, a SIC strategy is evoked to cancel and decode users' signals sharing the same pilot sequence.

Without loss of generality, we assume, in a two-UEs case $\left \| \textbf{h}_1 \right \| > \left \| \textbf{h}_2 \right \|$, 
on a SA basis, the BS conducts the SIC according to the descending order of channel gains, {\it i.e.,} the strongest UE-1 signal is decoded first without SIC application followed by UE-1 signal reconstruction and cancellation from the weakest UE-2 signal. 
Therefore, the signal-to-interference-plus-noise ratio (SINR) {on} each SA basis considering the two overlapping users $\gamma_1^{(b)}$ and $\gamma_2^{(b)}$ 
selecting the same pilot sequence $\tau_{\text{RA}}$ can be defined as:
%
%
\vspace{-3mm}
\begin{align} 
 \gamma_{1}^{(b)} = \frac{\rho_1 \beta_1^{(b)}}{\rho_2 \beta_2^{(b)} \sigma^2}, \qquad    
\gamma_{2}^{(b)} = \frac{\rho_2 \beta_2^{(b)}}{\varpi_1 (\rho_1 \beta^{(b)}_1) + \sigma^2}. 
\vspace{-6mm}
\label{eq:SINR} 
\end{align} 
where $\varpi_1$ is the residual interference factor after first step SIC
assuming imperfect signal reconstruction.

\noindent\textit{\textbf{Step IV}}: The BS estimate the channel to the user, or users, using the pilot signal sent in Step 3, and tries to decode the corresponding message \cite{Pereira2021}. If the decoding is successful, the BS has identified users in set $\mathcal{S}_t$ and admits its to the payload coherence blocks by allocating a dedicated data pilot sequence. If the decoding fails, the protocol has failed to resolve the collision and the unsuccessful UE is instructed to try again after a random interval, limited in a maximal of 10 attempts, then it stops sending pilot sequence and the packet is considered lost.

\subsection{System Sum-Rate} 
Defining the sum rate metric to compare systems. In SUCRe-XL only one user is accepted per pilot sequence in non-overlapping VR, on the other hand, in NVR-XL there will also be two users who consider themselves winners in the contender. The sum rate of NVR-XL on SA basis can be defined using \eqref{eq:SINR} as:
\vspace{-3mm}
\begin{equation}
R^{\text{NVR-XL}}_{\sum} = \sum_{b=1}^B \sum^{\tau_{\text{RA}}}_{t=1}  \sum_{k=1}^{\left | \mathcal{S}_t \right |} \log_2 \left ( 1+ \gamma^{(b)}_{t,k} \right ),  \quad [\rm bpcu]
\end{equation}
 The sum rate of SUCRe-XL can be defined as: 
\begin{equation}
    R_{\sum}^{\text{SUCRe-XL}} = \sum_{b \in \mathcal{V}_k} \sum^{\tau_{\text{RA}}}_{t=1} \log_2 \left ( 1 + \frac{\rho_k \beta_k^{(b)}}{\sigma^2} \right ), \quad [\rm bpcu]
\end{equation}
\noindent assuming that just one user per pilot sequence is accepted.
\vspace{-2mm}
\section{Numerical Results}

 
It is assumed a 100 meter URA with $M = 500$ antennas in a 200 x 100 $m^2$ cell with $K$ inactive users (iUES) randomly distributed  as illustrated in Fig. \ref{fig:system}, a number of available pilots $\tau_{\text{RA}} = 10$, and normalized transmit power $\rho_k = 1 \rm W$. The main simulation parameters is listed in Table \ref{tab:simulations}. {The scenario has simulated in Matlab 2019 utilizing one Intel HD Grafics 6000 GPU, Intel(R) Dual-Core(TM) I5 CPU @ 1.6 GHz and 8 GB RAM.}
\vspace{-3mm}
\begin{table}[ht]
\caption{Simulation Parameters.}
\begin{tabular}{l|c}
\hline \hline
\textbf{Parameter}           & \textbf{Value}      \\ 
\hline \hline
User distance to array  & $r_{k,m_z,m_y} \in {[}10;180\text{m}{]}$ \\ \hline
User length & $h_{user} \in {[}1;1.7 \text{m}{]}$ \\ \hline
Antennas elements y-axys, z-axys & $M_y = 100, M_z = 5 $ \\ \hline
Number of BS antennas (URA)  & $M = M_y \times M_z = 500$     \\ \hline
Antenna element distance & $d_m = 1$m \\ \hline
Array height   & $h_{array} = 12$m \\ \hline
Array length   & Ly = $M_y \times d_m$ \\ \hline 
User transmit power       &  $\rho_k = 1 \rm W$     \\ \hline
Noise Power     & $\sigma^2 = 1 \rm W$ \\ \hline
Number of iUEs               & $K \in {[}0;5000{]}$ \\ \hline
Access probability ($1^{st}$ attempt) & $P_a = 0.01$ \\ \hline
Access probability (new attempts)       & $P_{na} = 0.5$    \\ \hline
Number of available pilot sequences     & $\tau_{\text{RA}} = 10$     \\ \hline
SIC imperfection factor                 & $\varpi = 0.1$     \\ \hline
Monte Carlo realizations     & 5000 \\ \hline
\end{tabular}
\label{tab:simulations}
\vspace{-3mm}
\end{table}

\subsection{Average Number of Access Attempts} 
\begin{figure}[!htbp]
\centering
\includegraphics[width=.9\linewidth]{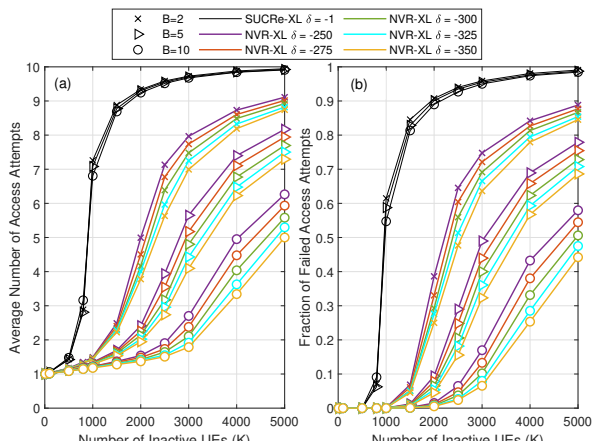}
\vspace{-5mm}
\caption{RA systems performance. (a)  Avg. Number of RA Attempts. (b) Prob. of failed access attempts.}
\label{fig:ANAA}
\vspace{-3mm}
\end{figure}

Fig. \ref{fig:ANAA} shows that the proposed protocol NVR-XL in crowded scenarios $(1000 < K < 5000)$  achieves improved RA performance. The modification in collision resolution is that NVR-XL accepts two users colliding on the overlapping VR, thanks to the power diversity; therefore, the superposition signal can be processed by SIC in each SA. With the increase in the number of subarrays $B$, but constrained by the channel hardening and favorable propagation properties, the probability of collision in each subarray by sharing overlapping VRs in each SA increases. Such collision resolution configuration is impossible to treat with SUCRe-XL due to the absence of the interference cancellation step. As a result, the NVR-XL allows decreasing the number of access attempts even under solid interference in the same SA.

\begin{figure}[!htbp]
\centering
\includegraphics[width=.89\linewidth]{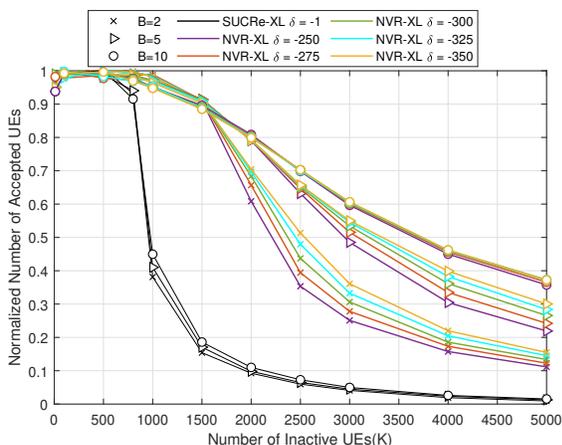} \vspace{-4mm}
\caption{Normalized number of accepted UEs.}
\label{fig:extra}
\end{figure}

The normalized number of accepted UEs, calculated by the number of users accepted by the number of users trying to access, given in Fig. \ref{fig:extra} corroborate with the improved results in NVR-XL to accept UEs to transmit due the superposition signal can be processed by SIC in each SA.  

\vspace{-6mm}
\subsection{Systems sum-rate} 
\begin{figure}[!htbp]
\centering
\includegraphics[width=.85\linewidth]{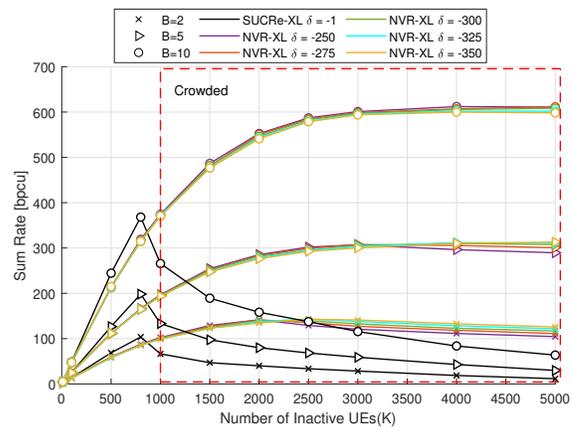}
\vspace{-4mm}
\caption{Avg sum-rate as function of number of iUEs.}
\label{fig:SumRate}
\vspace{-5mm}
\end{figure}

To adjust numerically the best value of scale factor, $\delta^{K,B^*}_{\textsc{nvr-xl}}$, we chose the scenario with $B = 10$ SAs and deploys an exhaustive search approach to find the best $\delta$ that maximizes the sum-rate; we have found the interval $\delta_{\textsc{nvr-xl}}^{(K,10)^*} = [-350:-250]$.
Fig. \ref{fig:SumRate} reveals the sum-rate performance for both protocols; the SUCRe-XL requires the non-overlapping VR em each SA to accept the device; hence, presents a modest gain even increasing the number of SA. The highest sum rate result occurs before the saturation of usage of available pilots. On the other hand, the NVR-XL processes the signal superimposed in each SA, canceling the more robust user's signal, with a residual interference in order of $\varpi_1 = 0.1$, and achieving improved results when $B$ increases. 
Also, a maximum sum rate limit is reached when a pair of users is processed in each SA in each RA attempt, as the SIC is able to cancel just one interfering signal, in the growing crowded scenario interference in each SA continues to increase (three or more users accessing the same SA with the same pilot).

\section{Conclusion}

This work proposes to explore the power domain to improve the performance of SUCRe-XL grant-based RA protocol; indeed, the XL-MIMO adds a new degree of freedom with the received signal being processed by different SAs, together with a new degree of freedom introduced by NOMA, allows to resolve the collision of two users using the same pilot sequence, with application of a SIC in each SA at the BS side. Our numerical results reveal a substantial performance improvement of the proposed RA protocol regarding the recent literature SUCRe-XL method, attaining simultaneously a decreasing of the number of access attempts and a substantial increment in the sum-rate for the crowded scenarios analyzed $(1000<K<5000)$.
\vspace{-5mm}
\bibliographystyle{IEEEtran}
\bibliography{ref1}

\end{document}